\pdfoutput=1
\documentclass[a4paper]{article} 
\usepackage{spconf,amsmath,graphicx} 
\newcommand{\bm}[1]{\mbox{\boldmath$#1$}} 

\title{Fast and High-Quality Singing Voice Synthesis System \\based on Convolutional Neural Networks}
\name{
\begin{tabular}{c}
Kazuhiro Nakamura$^{\star}$, Shinji Takaki$^{\star \dagger}$, Kei Hashimoto$^{\star \dagger}$, Keiichiro Oura$^{\star \dagger}$, Yoshihiko Nankaku$^{\dagger}$,\\
 and Keiichi Tokuda$^{\star \dagger}$
\end{tabular}
}
\address{
$^{\star} $Department of Research and Development, Techno-Speech, Inc., Nagoya, Japan\\
$^{\dagger} $Department of Computer Science, Nagoya Institute of Technology, Nagoya, Japan
}

\begin{document}
\ninept
\maketitle

\vspace{-1mm}
\begin{abstract}
\vspace{-1mm}
The present paper describes singing voice synthesis based on convolutional neural networks ({CNN}s).
Singing voice synthesis systems based on deep neural networks ({DNN}s) are currently being proposed
and are improving the naturalness of synthesized singing voices.
As singing voices represent a rich form of expression, a powerful technique to model them accurately is required.
In the proposed technique, long-term dependencies of singing voices are modeled by {CNN}s.
An acoustic feature sequence is generated for each segment that consists of long-term frames,
and a natural trajectory is obtained without the parameter generation algorithm.
Furthermore, a computational complexity reduction technique, which drives the {DNN}s in different time units depending on type of musical score features, is proposed.
Experimental results show that the proposed method can synthesize natural sounding singing voices much faster than the conventional method.
\end{abstract}
\vspace{-2mm}
\section{Introduction}
\vspace{-2mm}
\label{sec:intro}

Deep neural networks ({DNN}s), which are artificial neural networks with many hidden layers,
are attaining significant improvement in various speech processing areas, e.g.,
speech recognition \cite{hinton-IEEE-2012}, speech synthesis \cite{zen-ICASSP-2013, qian-ICASSP-2014}
and singing voice synthesis \cite{nishimura-Interspeech-2016}.
In {DNN}-based singing voice synthesis, a {DNN} works as an acoustic model that represents a mapping function from musical score feature sequences
(e.g., phonetic, note key, and note length) to acoustic feature sequences (e.g., spectrum, excitation, and vibrato).
{DNN}-based acoustic models can represent complex dependencies
between musical score feature sequences and acoustic feature sequences more efficiently than hidden Markov model ({HMM})-based acoustic models \cite{watts-ICASSP-2016}.
Neural networks that can model audio waveforms directly, e.g., {W}ave{N}et \cite{oord-arXiv-2016},
{S}ample{RNN} \cite{mehri-arXiv-2016}, {W}ave{RNN} \cite{kalchbrenner-arXiv-2018},
{FFTN}et \cite{jin-ICASSP-2018}, and {W}ave{G}low \cite{prenger-arXiv-2018}, are currently being proposed.
Such neural networks are used as vocoders in the speech field and improve the quality of synthesized speech compared to conventional vocoders \cite{tamamori-ICASSP-2017}.
The neural vocoders use acoustic features as inputs.
Therefore, accurately predicting them from musical score features by using acoustic models
is still an important issue for generating high-quality speech or singing voices.

One limitation of the feed-forward {DNN}-based acoustic modeling \cite{zen-ICASSP-2013} is that the sequential nature of speech is not considered.
Although there certainly are correlations between consecutive frames in speech data,
the feed-forward {DNN}-based approach assumes that each frame is generated independently.
As a solution, recurrent neural networks ({RNN}s) \cite{robinson-NIPS-1988},
especially long short-term memory ({LSTM})-{RNN}s \cite{hochreiter-NeuralComput-1997},
provide an elegant way to model speech-like sequential data that embody short- and long-term correlations.
Furthermore, this problem can be mitigated by smoothing predicted acoustic features
with a speech parameter generation algorithm \cite{tokuda-ICASSP-2000}
that utilizes dynamic features as constraints to generate smooth speech parameter trajectories.
On the other hand, some techniques for incorporating the sequential nature of speech data into the acoustic model itself have been proposed \cite{zen-ICASSP-2015, wang-ICASSP-2017}.

This paper proposes an architecture that uses {CNN}s to convert musical score feature sequences into acoustic feature sequences segment-by-segment.
The proposed approach can capture long-term dependencies of singing voices and can generate natural trajectories 
without the use of the speech parameter generation algorithm \cite{tokuda-ICASSP-2000}.
The training of {CNN}s and the generation of acoustic features are fast, 
because there is no recurrent structure in this architecture.
Furthermore, an efficient technique to reduce the computational complexity during the generation of acoustic features is proposed.

The rest of this paper is organized as follows.
Related work is described in Section~2.
Details of the proposed {CNN}-based singing voice synthesis architecture and the computational complexity reduction technique are described in Section~3.
Experimental results are given in Section~4.
The key points are summarized, and future work is mentioned in Section~5.

\vspace{-2mm}
\section{Related work}
\vspace{-2mm}

\subsection{{DNN}-based singing voice synthesis}
\label{sec:dnn-based_singing_voice_synthesis}
In recent years, several kinds of {DNN}-based singing voice synthesis systems \cite{nishimura-Interspeech-2016, hono-APSIPA-2018,merlijn-arXiv-2017, yuan-interspeech-2019, juheon-interspeech-2019} have been proposed.
In the training part of the basic system \cite{nishimura-Interspeech-2016}, parameters for spectrum (e.g., mel-cepstral coefficients), excitation,
and vibrato are extracted from a singing voice database as acoustic features.
Vibrato is a singing expression wherein the pitch is periodically shaken.
Musical score feature sequences and acoustic feature sequences are then time-aligned by well-trained {HMM}s,
and the mapping function between them is modeled by {DNN}s.
The start timing of the singing voice is often earlier than that of the corresponding note.
Time-lag models have been introduced \cite{hono-APSIPA-2018} to predict such differences.
Furthermore, a musical-note-level pitch normalization technique has been proposed to generate various singing voices including arbitrary pitch \cite{nishimura-Interspeech-2016}.
In this technique, the differences between the log $F_0$ sequences extracted from waveforms and the pitch of musical notes are modeled.
In the synthesis part, an arbitrarily musical score that includes lyrics to be synthesized is first converted into a musical score feature sequence, and is then mapped to an acoustic feature sequence by the trained {DNN}s.
Next, the speech parameters (spectrum, excitation, and vibrato) are generated
by a maximum likelihood parameter generation ({MLPG}) algorithm \cite{tokuda-ICASSP-2000}.
It was shown that the quality of the generated speech was improved
by considering the explicit relationship between static and dynamic features \cite{hashimoto-ICASSP-2015}.
Finally, a singing voice is synthesized from the generated parameters
by using a vocoder based on a mel log spectrum approximation ({MLSA}) filter \cite{imai-IEICE-1983}.

\vspace{-2mm}
\subsection{Modeling long-term dependencies of speech}
\vspace{-1mm}
The simplest way to apply neural networks to statistical parametric speech synthesis ({SPSS}) \cite{zen-SCommu-2009} is to use a feed-forward neural network (FFNN) \cite{zen-ICASSP-2013}
as a deep regression model to map linguistic features directly to acoustic features.
One limitation of this architecture is the one-to-one mapping between linguistic and acoustic features.
{RNN}s \cite{robinson-NIPS-1988} provide an elegant way to model speech-like sequential data
that embody correlations between neighboring frames.
That is, previous input features can be used to predict output features at each frame.
{LSTM-RNN}s \cite{hochreiter-NeuralComput-1997}, which can capture long-term dependencies,
have been applied to acoustic modeling for {SPSS}.
Fan {\it et al.} and Fernandez {\it et al.} applied deep bidirectional {LSTM-RNN}s,
which can access input features at both past and future frames,
to acoustic modeling for SPSS and reported improved naturalness \cite{fan-Interspeech-2014, fernandez-Interspeech-2014}.
Trajectory training is another approach for capturing long-term dependencies of speech.
In {DNN}-based systems, although the frame-level objective function is usually used for {DNN} training,
the sequence-level objective function is used for parameter generation.
To address this inconsistency between training and synthesis, a trajectory training method was introduced into the training process of {DNN}s \cite{hashimoto-ICASSP-2016}.
The method was also applied to a singing voice synthesis framework \cite{hono-APSIPA-2018}.

The problem with the {RNN}-based systems is that they take time due to the difficulty of parallelizing model training and parameter generation.
And the problem with the trajectory training method is that the computational cost increases significantly as the sequence length increases.

\vspace{-2mm}
\subsection{Considering the sequential nature of acoustic features}
\vspace{-1mm}
One limitation of the {DNN}-based acoustic modeling is that the sequential nature of acoustic features is not sufficiently taken into consideration.
Although this problem can be mitigated by smoothing the predicted acoustic features using the speech parameter generation algorithm \cite{tokuda-ICASSP-2000},
which utilizes dynamic features as constraints to generate smooth trajectories.
However, as many text-to-speech ({TTS}) applications require fast and low-latency speech synthesis,
an existing problem is the high-latency that the {MLPG} algorithm causes during generation.
Fan {\it et al.} claimed that deep bidirectional {LSTM-RNN}s can generate smooth speech parameter trajectories; thus, no smoothing step was required \cite{fan-Interspeech-2014},
whereas Zen {\it et al.} reported that having the smoothing step was still necessary with unidirectional {LSTM-RNN}s \cite{zen-ICASSP-2013}.

An efficient way to solve this problem is to incorporate the sequential nature of speech data into the acoustic model itself.
Zen {\it et al.} proposed a recurrent output layer \cite{zen-ICASSP-2015},
whereas Wang {\it et al.} proposed a convolutional output layer \cite{wang-ICASSP-2017},
to achieve smooth transitions between consecutive frames, and accordingly, to replace the {MLPG}.
They were used with unidirectional {LSTM} to achieve both natural sounding speech and low-latency speech synthesis.

\vspace{-2mm}
\section{{CNN}-based singing voice synthesis}
\vspace{-2mm}

\begin{figure}[t]
\centering
  \includegraphics[width=\linewidth]{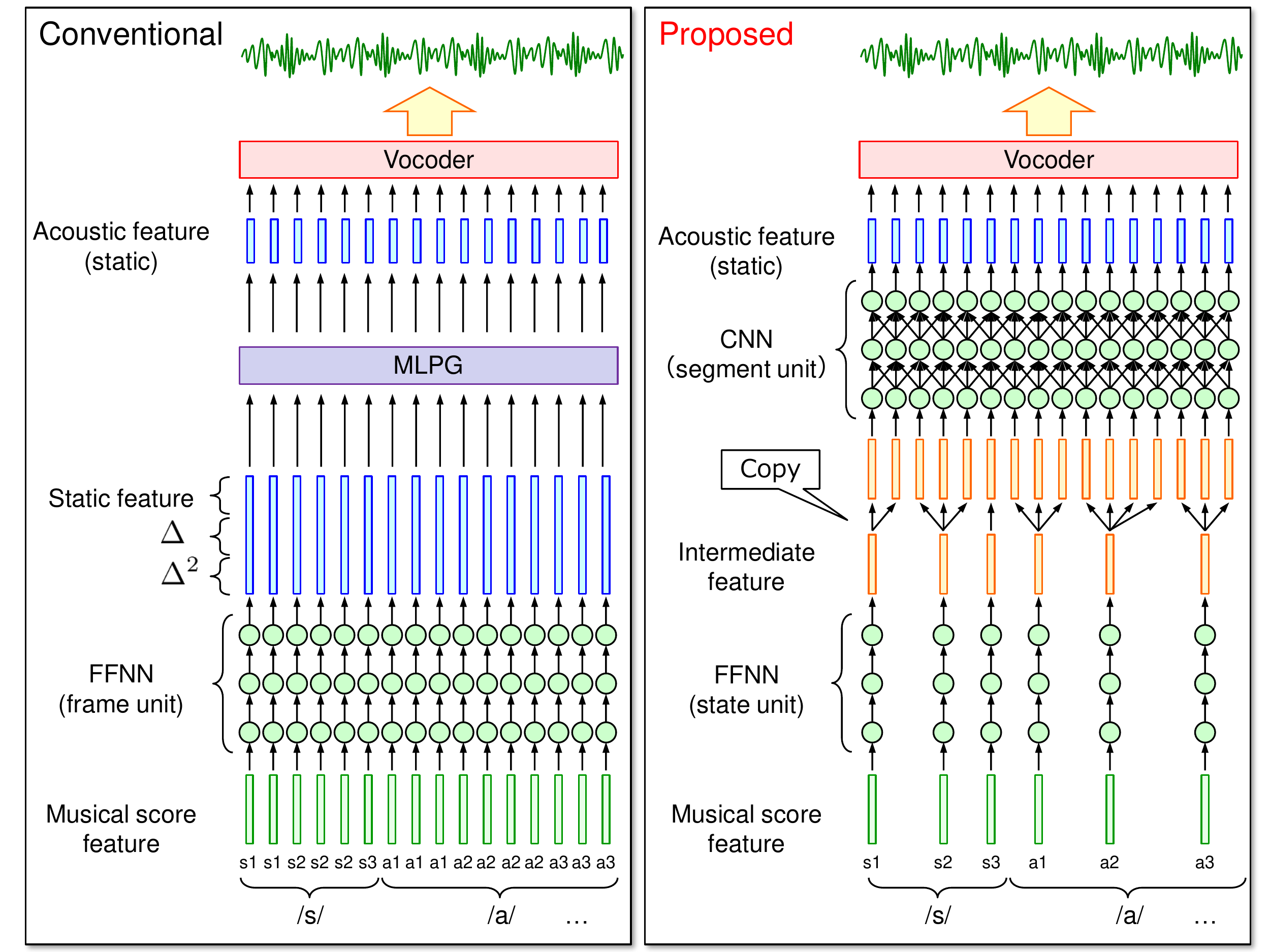}
  \vspace{-8mm}
\caption{Difference between conventional and proposed methods.}
  \label{fig:method_comparison}
  \vspace{-4mm}
\end{figure}

\subsection{{CNN}-based architecture for capturing long-term dependencies of singing voice}
\vspace{-1mm}
In the proposed method, the relatively long musical score feature sequence, equivalent to several seconds to tens of seconds,
is regarded as one segment and converted into the acoustic feature sequence by {CNNs} at the same time.
The difference between the conventional \cite{nishimura-Interspeech-2016} and proposed methods is shown in Figure~\ref{fig:method_comparison}.
In the proposed method, the first part consists of $1 \times 1$ convolutional layers
equivalent to {FFNNs} and converts the musical score feature sequence one-by-one.
The dropout technique is used to keep the robustness against the unknown musical scores.
The second part consists of $1 \times n$ convolutional layers,
where the intermediate output feature sequence of the first part is converted
into the acoustic feature sequence segment-by-segment.
The dimension of the acoustic features is expressed as the number of channels of the output features.
The size of the segment is $1 \times T$, where $T$ means the number of frames in each segment.
Since a fully convolutional network ({FCN}) \cite{long-CVPR-2015} is used as the {CNN} structure, the segment size $T$ is adjustable.
These two parts are integrated and trained simultaneously.

The relationship between a musical score feature vector sequence $\bm{s}=[\bm{s}_1^{\top}, \bm{s}_2^{\top}, \dots, \bm{s}_T^{\top}]^{\top}$
and an acoustic feature vector sequence $\bm{c}=[\bm{c}_1^{\top}, \bm{c}_2^{\top}, \dots, \bm{c}_T^{\top}]^{\top}$
is represented as follows:
\vspace{-2mm}                                                                                                                                                                                                                       
\begin{eqnarray}
\bm{c} = G([F(\bm{s}_1)^{\top}, F(\bm{s}_2)^{\top}, \dots, F(\bm{s}_T)^{\top}]^{\top}),
\end{eqnarray}
where $F(\cdot)$ is an one-by-one mapping function in the first part,
and $G(\cdot)$ is a segment-by-segment mapping function in the second part.

\vspace{-1mm}  
\subsection{Loss function for obtaining smooth parameter sequence without use of the parameter generation algorithm}
\vspace{-1mm}  
In the proposed method, a loss function based on the likelihood of $\bm{o}_t$ is used to obtain smooth acoustic feature sequences.
A parameter vector $\bm{o}_t$ of a singing voice consists of a $D$-dimensional static feature vector
$\bm{c}_t = \left[c_t (1), c_t (2), \dots, c_t (D) \right]^\top$
and their dynamic feature vectors $\Delta^{(\cdot)} \bm{c}_t$.
\vspace{-1mm}
\begin{eqnarray}
\bm{o}_t = [\bm{c}_t^\top, \Delta^{(1)} \bm{c}_t^\top, \Delta^{(2)} \bm{c}_t^\top ]^\top
\end{eqnarray}
The sequences of the singing voice parameter vectors and the static feature vectors can be written in vector forms as follows:
\begin{eqnarray}
\bm{o} &=& [\bm{o}_t^\top, \dots, \bm{o}_t, \dots, \bm{o}_T^\top]^\top \\
\bm{c} &=& [\bm{c}_t^\top, \dots, \bm{c}_t, \dots, \bm{c}_T^\top]^\top,
\end{eqnarray}
where $T$ is the number of frames.
The relation between $\bm{o}$ and $\bm{c}$ can be represented by $\bm{o} = \bm{W}\bm{c}$,
where $\bm{W}$ is a window matrix extending the static feature vector sequence $\bm{c}$
to the singing voice parameter vector sequence $\bm{o}$.

In the training part, an objective function is defined as
\begin{eqnarray}
\mathcal{L} = \mathcal{N}\left(\bar{\bm{o}} | \bm{o} , \bm{\Sigma}\right),
\end{eqnarray}
where $\bar{\bm{o}}$ is represented by $\bar{\bm{o}} = \bm{W}\bar{\bm{c}}$,
where $\bar{\bm{c}}$ is the static feature vector sequence of the recorded singing voice.
$\bm{\Sigma}$ is a globally tied covariance matrix given by
\begin{eqnarray}
\bm{\Sigma} = diag [\bm{\Sigma}_1, \dots, \bm{\Sigma}_t, \dots, \bm{\Sigma}_T] 
\end{eqnarray}
and is updated during the training.

The proposed method can generate a natural trajectory without the parameter generation algorithm
by considering not only static features but also dynamic features in the training part of the {CNNs}.

\vspace{-1mm} 
\subsection{Computational Complexity Reduction}
\vspace{-1mm} 
As the DNN-based singing voice synthesis systems generally require high computational complexity,
we propose a computational complexity reduction technique that keeps the naturalness of the synthesized singing voices.
Although the music score feature parameters in song, phrase, note, syllable, phoneme, state, and frame levels have been treated the same way,
it is efficient in terms of computational complexity to input them in stages according to the temporal resolution of features.
Namely, the frame level features should be processed in each frame, 
whereas the song level features should be processed only once.
The right part of Figure~\ref{fig:method_comparison} shows the proposed methods.
In this method, the features obtained from the musical score and the state number are converted state-by-state by FFNNs,
expanded to the frame level, concatenated to the position parameters in frame level,
and then converted segment-by-segment into acoustic features by CNNs.
As the result, the FFNNs is driven once for each state, making it possible to greatly reduce the computational complexity.
In addition, since the acoustic feature sequence is generated in units of segments that consist of long-term frames,
it is possible to synthesize natural singing voices. 

In recent years, as such a network having different driving levels,
there is an attention mechanism which has been used in end-to-end speech synthesis systems \cite{shen-icassp-2018, li-arXiv-2018}.
In this approach, features in the frame level are generated by weighted sum of intermediate features in the phoneme level and used as the input to the decoder module.
Although it is desirable to use such a mechanism in singing voice synthesis,
the attention mechanism requires a large amount of training data and increases the computational complexity during synthesis.
In addition, as the note timing in the musical scores have to be considered,
the attention mechanism used in text synthesis cannot be applied as it is.
Thus, in the proposed method, a single {V}iterbi path estimated by well-trained {HMM}s and the frame-level-position parameters obtained from the state boundaries are used in place of the attention.

\vspace{-2mm} 
\section{Experiments}
\vspace{-2mm} 
\subsection{Experimental conditions}
\vspace{-1mm} 
Two tests were conducted to evaluate the effectiveness of the proposed method.
The first test is an evaluation of the quality of the synthesized singing voices for the conventional and proposed methods.
A conventional vocoder and a WaveNet vocoder were used for conversion from acoustic feature sequences to singing voice waveforms ({TEST}1). 
The other test is an evaluation of the relationship between computational complexity and quality ({TEST}2).

For the training, 55 Japanese children's songs and 55 {J-POP} songs by a female singer were used,
and for the test, 5 other {J-POP} songs were used.
The test data were divided into phrases, whose average length was 9.8 seconds.
Singing voice signals were sampled at 48 {kHz} and windowed with a 5-ms shift.
The number of quantization bits was 16.
The feature vectors consisted of 0-th through 49-th STRAIGHT \cite{kawahara-SpeechCommunication-1999} mel-cepstral coefficients,
log ${F_0}$ values, 22-dimension aperiodicity measures, and 2-dimension vibrato parameters.
The vibrato parameter vectors consisted of amplitude (cent) and frequency ({Hz}).
The areas that do not have a value were interpolated linearly for log ${F_0}$ and vibrato parameters,
and two binary flags representing voiced / unvoiced and with / without vibrato are included.
The input features that include 724 binary features for categorical contexts (e.g., the current phoneme identity and the key of the current measure)
and 122 numerical features for numerical contexts (e.g., the number of phonemes in the current syllable and the absolute pitch of the current musical note) were used.
Both input and output features in the training data for the {DNN}s were normalized;
the input features were normalized to be within 0.00--1.00,
and the output features were normalized to be within 0.01--0.99 on the basis of their minimum and maximum values in the training data.

As the pitch of musical notes greatly affects the synthesized singing voices,
the input features of DNNs include log $F0$ parameters of musical notes.
In particular, the alignment of musical notes is adjusted in accordance with the recorded singing voices,
and the log $F_0$ parameter sequence which represents the pitch of musical notes is used.
The areas of the log $F_0$ parameter sequence corresponding to the musical rests in musical scores are interpolated linearly.
The effectiveness was confirmed in the preliminary subjective experiment.

The FFNN-based singing voice synthesis system \cite{nishimura-Interspeech-2016} was used as the conventional method.
The conventional system had 3 hidden layers with 2048 units, and dropout with probability of 0.2 was used.
The acoustic features and their dynamic features (delta and delta-delta) were the output features,
and the {MLPG} algorithm was used to obtain the smooth feature sequences.
In the proposed system, the first part consisted of 3 hidden layers of FFNNs.
The second part had 2 layers for down-sampling, multiple layers that have residual structure, and 2 layers for up-sampling.
In these layers, the filter size was 3, and the stride of the down- and up-sampling layers was 2.
The data were separated into segments of 2000 frames and used for training and generation,
and 100 adjacent frames were cross-faded in the generation step.
In both systems, the ReLU activation function was used for hidden layers, and the sigmoid function was used for the output layer.
Five-state, left-to-right, no-skip, hidden semi-Markov models ({HSMMs}) were used to obtain the time alignment of acoustic features in state for training,
and the state durations of the test songs were predicted by {FFNNs} trained from the time alignment of the training data.

\vspace{-1mm}
\subsection{Experimental results of TEST1}
\vspace{-1mm}
\label{sec:subjective}

\begin{figure}[t]
  \centering
    \vspace{-2mm}
  \includegraphics[width=1.0\linewidth]{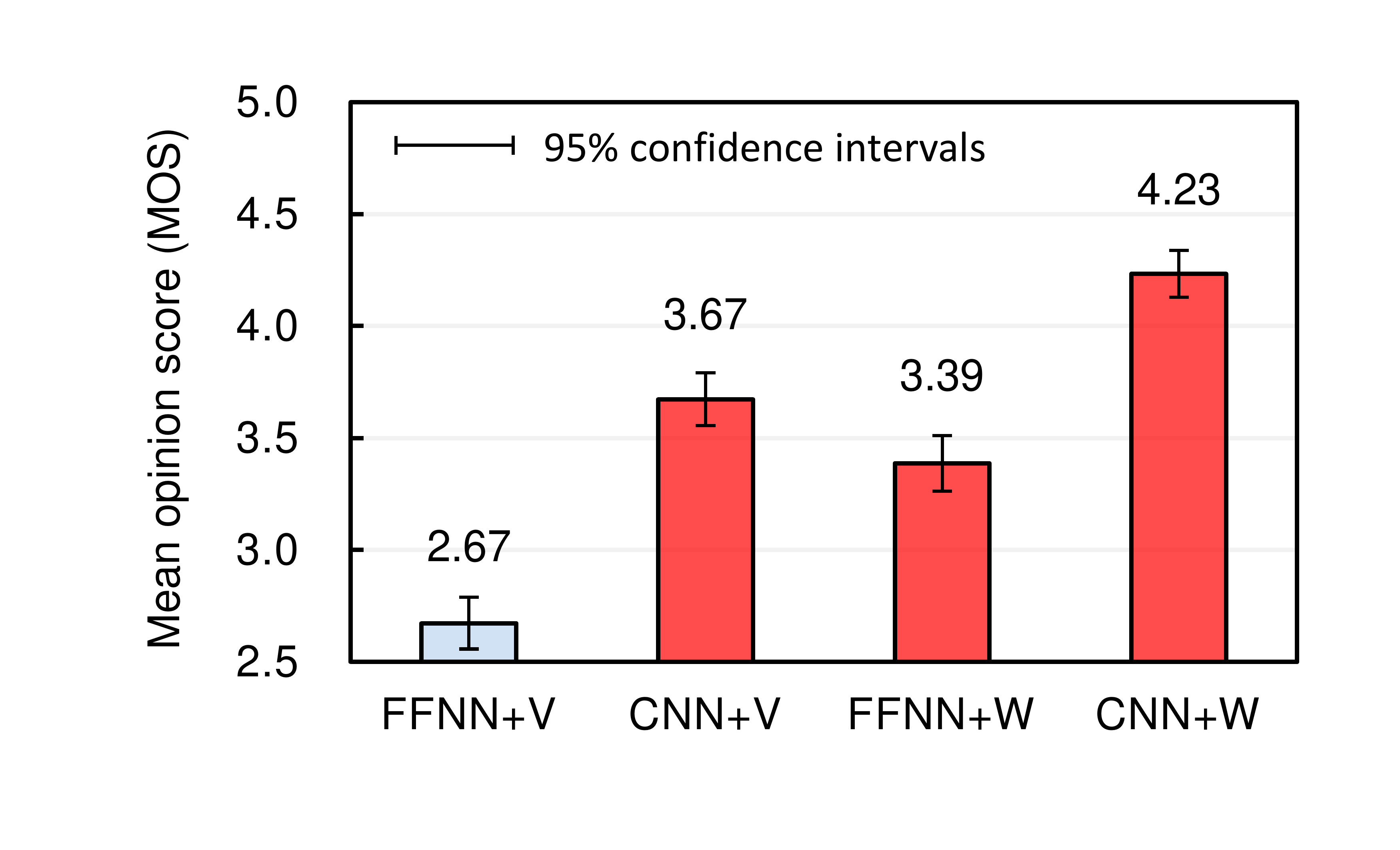}
   \vspace{-13mm}
  \caption{Subjective experimental results of TEST1.}
  \vspace{-7mm}
  \label{fig:mos_vocoder_wavenet}
\end{figure}
In {TEST}1, a subjective comparison test of mean opinion scores ({MOS}) was conducted.
The computational complexity reduction technique was not used in this test.
A {MLSA}-based vocoder \cite{imai-IEICE-1983} and a {W}ave{N}et vocoder \cite{tamamori-ICASSP-2017} were used
for conversion from acoustic feature sequences to singing voice waveforms.
The singing voice signals to train {W}ave{N}et were sampled at 48 kHz and quantized from 16 bits to 8 bits by using the $\mu$-law quantizer \cite{PCM}.
Mel-cepstrum-based noise shaping and prefiltering were applied to the quantization step \cite{yoshimura-IEEE-2018}.
The parameters for adjusting the intensity in the noise shaping and prefiltering were set as $\gamma=0.4$, $\beta=0.2$.
The dilations of the {W}ave{N}et model were set to $1, 2, 4, \dots, 512$.
Ten dilation layers were stacked three times.                          
The sizes of the channels for dilations, residual blocks, and skip-connections were 256, 512, and 256, respectively.
The 5-point {MOS} evaluation for naturalness was conducted.
Fifteen subjects evaluated ten phrases that were randomly selected from the test data for each method.

Figure~\ref{fig:mos_vocoder_wavenet} shows the results of the MOS evaluation.
\textbf{FFNN+V} and \textbf{FFNN+W} represent conventional systems using the {MLPG} algorithm,
and \textbf{CNN+V} and \textbf{CNN+W} represent proposed systems.
\textbf{V} and \textbf{W} mean {MLSA}-based vocoder and {W}ave{N}et vocoder, respectively.

The proposed systems (\textbf{CNN+V} and \textbf{CNN+W})
outperformed the conventional {FFNN}-based systems (\textbf{FFNN+V} and \textbf{FFNN+W})
as shown in Figure~\ref{fig:mos_vocoder_wavenet}.
These results indicate that the naturalness of the synthesized singing voice is drastically improved by modeling the time-dependent variation with {CNN}s.
Moreover the {W}ave{N}et vocoder (\textbf{FFNN+W} and \textbf{CNN+W}) showed a better score than the {MLSA}-based vocoder
(\textbf{FFNN+V} and \textbf{CNN+V}), respectively.
Some samples of \textbf{CNN+W} are available for listening \cite{techno-speech-web}.

\begin{figure}[t]
  \centering
  \includegraphics[width=1.0\linewidth]{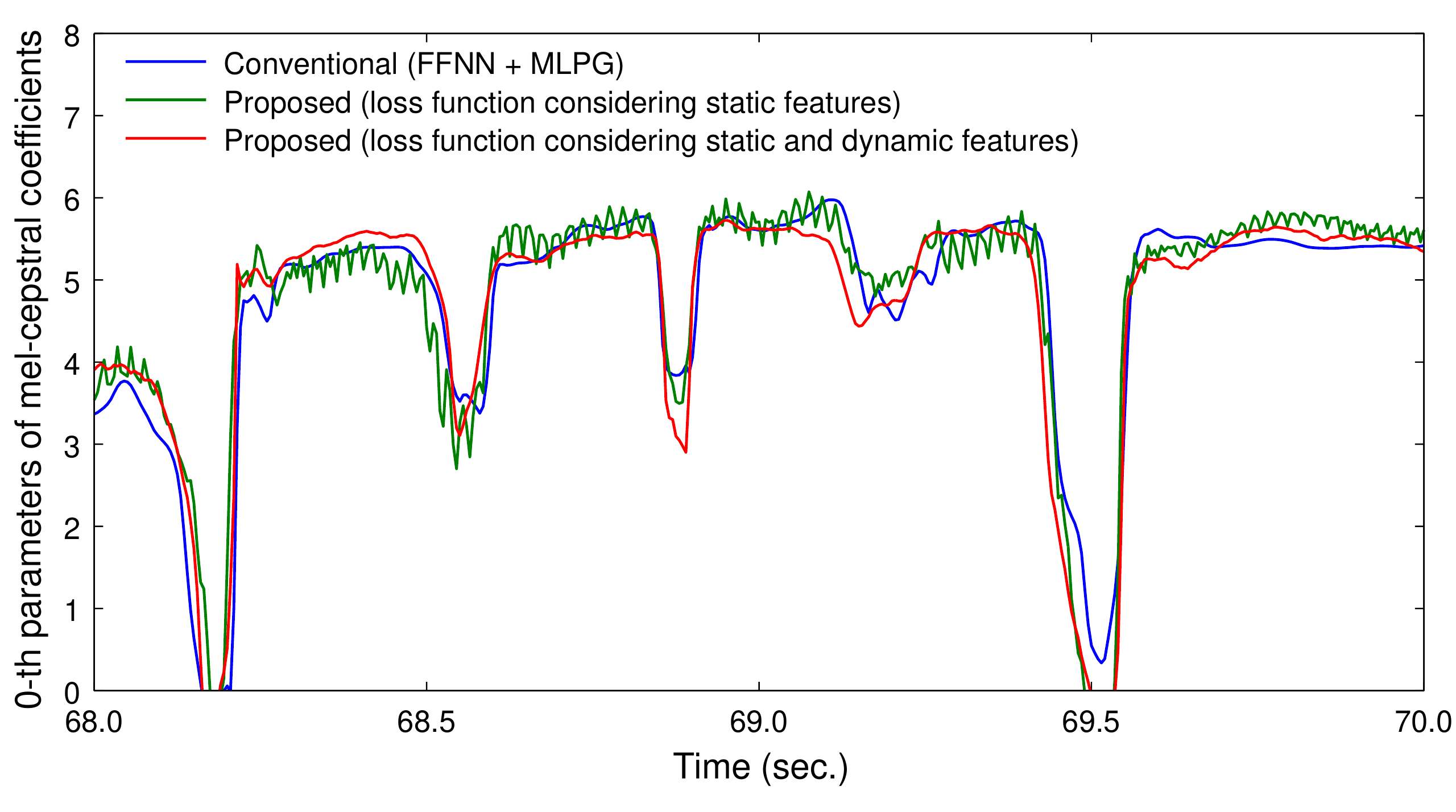}
    \vspace{-9mm}
  \caption{Comparison of 0-th parameters of mel-cepstral coefficients.}
  \vspace{-4mm}
  \label{fig:c0_sequence}
\end{figure}

An example of the generated parameter sequences is shown in Figure~\ref{fig:c0_sequence}.
Comparison of the two proposed methods shows that considering the loss of the dynamic features is effective to obtain a smooth parameter sequence.

\vspace{-1mm}
\subsection{Experimental results of TEST2}
\vspace{-1mm}
In {TEST}2, computational complexity was measured, and {MOS} evaluation was conducted.
A {MLSA}-based vocoder was used in all methods.
One thread of {I}ntel {C}ore i7-6700 {CPU} was used for measuring time in each method.
The 5-point {MOS} evaluation for naturalness was conducted.
Sixteen subjects evaluated ten phrases that were randomly selected from the test data for each method.

\begin{figure}[t]
  \centering
  \vspace{-2mm}
  \includegraphics[width=1.0\linewidth]{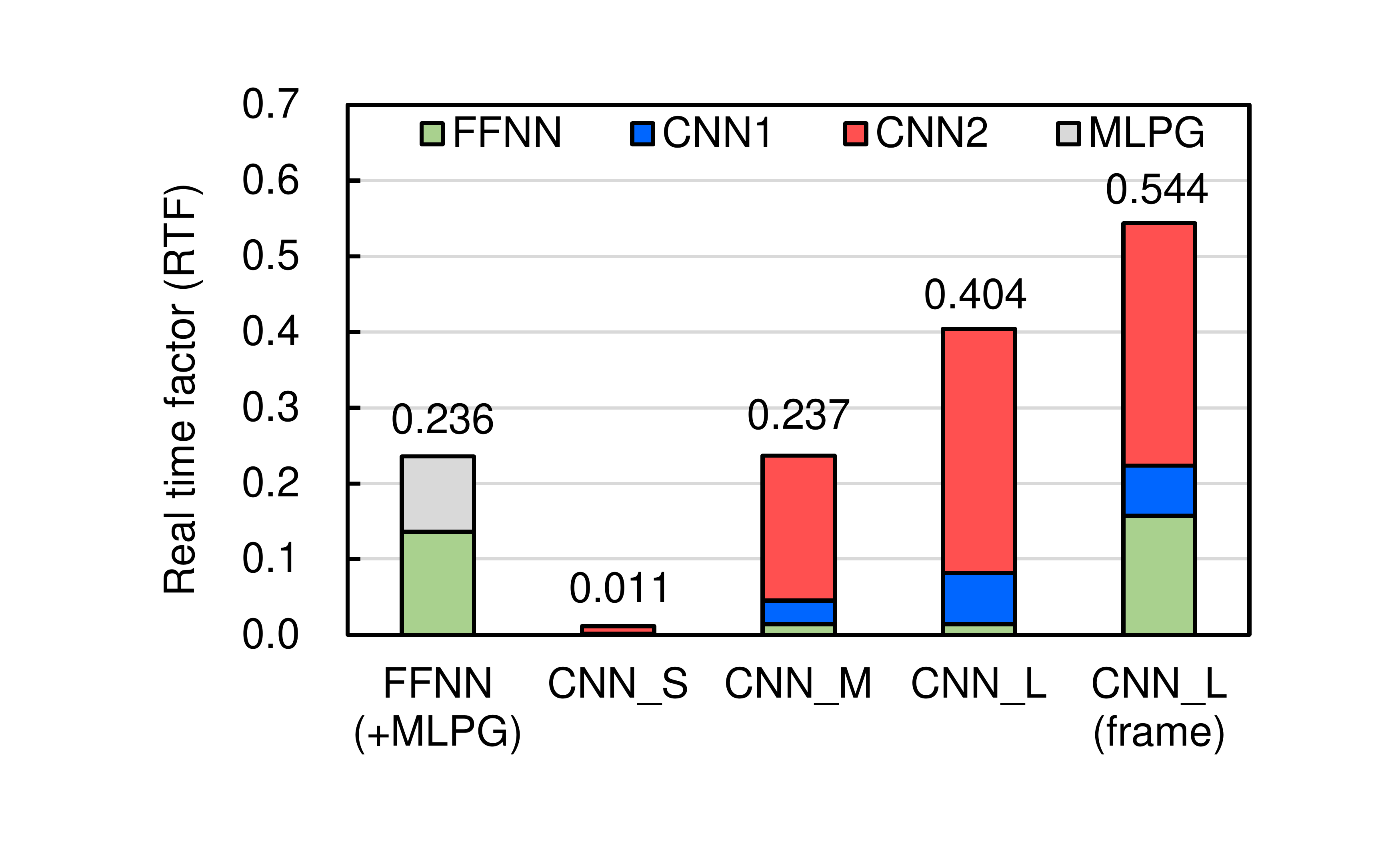}
  \vspace{-12mm}
  \caption{Computational times in TEST2.}
  \vspace{-4mm}
  \label{fig:fast_speed}
\end{figure}

\begin{figure}[t]
  \centering
  \vspace{-2mm}
  \includegraphics[width=1.0\linewidth]{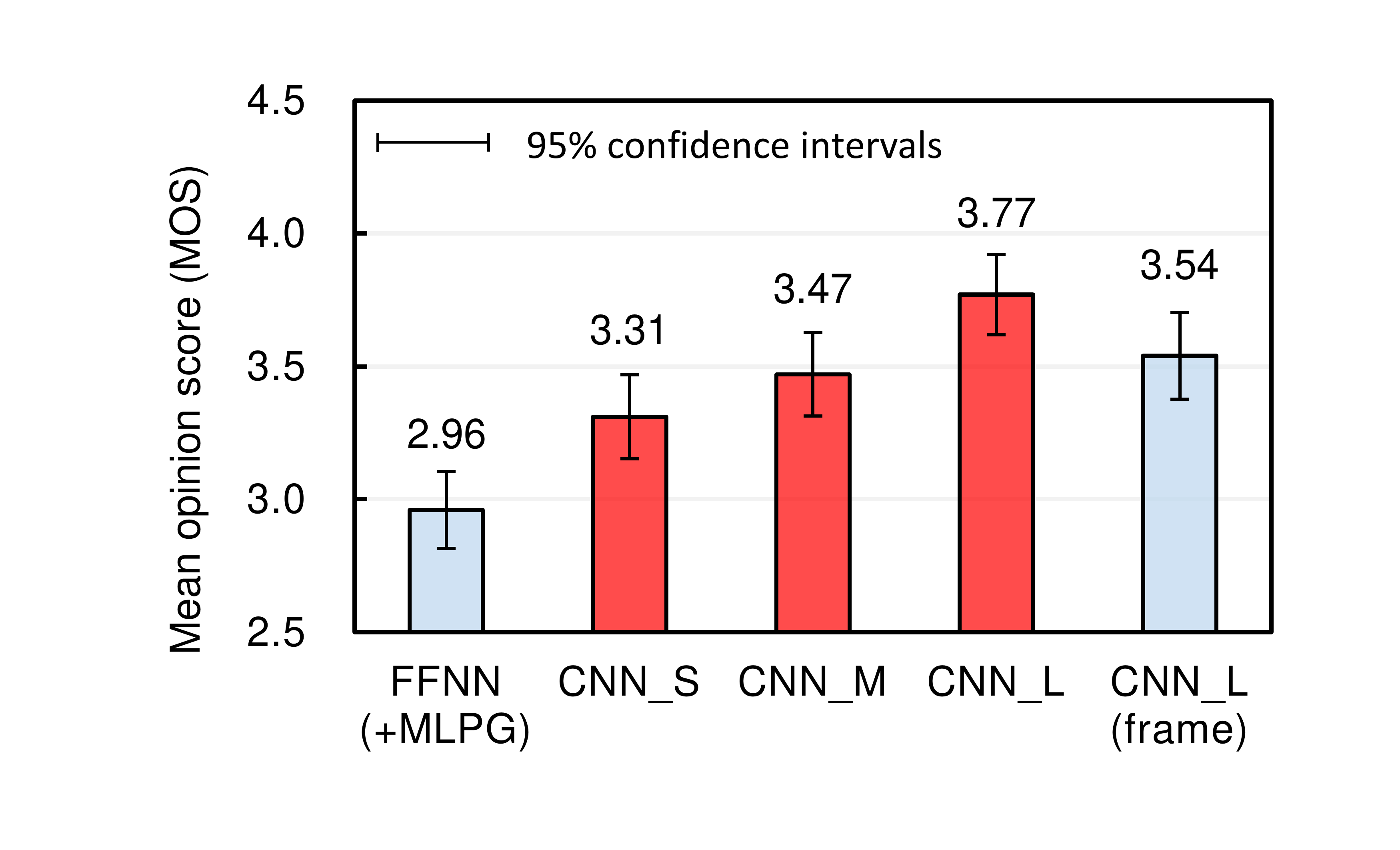}
    \vspace{-12mm}
  \caption{Subjective experimental results of TEST2.}
  \vspace{-4mm}
  \label{fig:fast_mos}
\end{figure}

Figure~\ref{fig:fast_speed} and Figure~\ref{fig:fast_mos} show the results of computational complexity measurement and the MOS evaluation, respectively.
\textbf{FFNN(+MLPG)} represents the conventional system.
The computational complexity reduction technique was used for \textbf{CNN\_S}, \textbf{CNN\_M}, and \textbf{CNN\_L},
and not used for \textbf{CNN\_L(frame)}.
The model sizes of \textbf{CNN\_S} and \textbf{CNN\_M} were adjusted so that the computational time was about 5\% and 100\%, respectively, of the conventional method.
The model sizes of \textbf{CNN\_L} and \textbf{CNN\_L(frame)} were the same as \textbf{CNN+V} in {TEST}1.
In the proposed methods, the numbers of residual layers were 5 for \textbf{CNN\_S}, and 9 for the others.
Except for \textbf{CNN\_S}, the {CNN} part of the proposed method was divided into a small {CNN} ({CNN}1) that outputs editable parameters (c0, log $F0$, vibrato amplitude, and frequency) 
and a large {CNN} ({CNN}2) that outputs the other parameters to draw the editable parameters immediately in {GUI} applications.

\textbf{CNN\_S} improved its naturalness even though its computational complexity was reduced to about 5\% compared to \textbf{FFNN}.
Compared to models of the same size without the computational complexity reduction technique,
the computational times for \textbf{CNN\_S}, \textbf{CNN\_M}, and \textbf{CNN\_L} were reduced by about 54\%, 38\%, and 26\%, respectively.
Comparison of \textbf{CNN\_L} and \textbf{CNN\_L(frame)} shows that the naturalness was not degraded by the computational complexity reduction technique.

\vspace{-2mm}
\section{Conclusions}
\vspace{-2mm}
In this paper, we proposed a {CNN}-based acoustic modeling technique for singing voice synthesis.
Long-term dependencies of singing voices that contain rich vocal expressions were modeled by {CNN}s.
Musical score feature sequences from musical scores were converted into acoustic feature sequences segment-by-segment,
and natural speech parameter trajectories were obtained without using the conventional speech parameter generation algorithm.
We also described a computational complexity reduction technique.
Experimental results showed that the proposed system generates more natural synthesized singing voices,
and can reduce computational complexity without degradation of the naturalness.

Future work includes comparison with {RNN}-based methods, 
evaluation of the proposed architecture on {TTS},
and tuning of parameters for practical use.
\vspace{-2mm}
\section{Acknowledgements}
\vspace{-2mm}
This work was supported by the MIC/SCOPE \#182206001.

\vfill\pagebreak

\bibliographystyle{main}
\bibliography{main}

\begin{thebibliography}{10}

\bibitem{hinton-IEEE-2012}
G.~Hinton, L.~Deng, D.~Yu, G.~E. Dahl, A.~Mohamed, N.~Jaitly, A.~Senior,
  V.~Vanhoucke, P.~Nguyen, T.~Sainath, and B.~Kingsbury,
\newblock ``Deep neural networks for acoustic modeling in speech recognition:
  The shared views of four research groups,''
\newblock {\em IEEE Signal Processing Magazine}, vol. 29, no. 6, pp. 82--97,
  2012.

\bibitem{zen-ICASSP-2013}
H.~Zen, A.~Senior, and M.~Schuster,
\newblock ``Statistical parametric speech synthesis using deep neural
  networks,''
\newblock {\em Proceedings of ICASSP}, pp. 7962--7966, 2013.

\bibitem{qian-ICASSP-2014}
Y.~Qian, Y.~Fan, W.~Hu, and F.~K. Soong,
\newblock ``On the training aspects of deep neural network ({DNN}) for
  parametric {TTS} synthesis,''
\newblock {\em Proceedings of ICASSP}, pp. 3829--3833, 2014.

\bibitem{nishimura-Interspeech-2016}
M.~Nishimura, K.~Hashimoto, K.~Oura, Y.~Nankaku, and K.~Tokuda,
\newblock ``Singing voice synthesis based on deep neural networks,''
\newblock {\em Proceedings of Interspeech}, pp. 2478--2482, 2016.

\bibitem{watts-ICASSP-2016}
O.~Watts, G.~E. Henter, T.~Merritt, Z.~Wu, and S.~King,
\newblock ``From {HMM}s to {DNN}s: where do the improvements come from?,''
\newblock {\em Proceedings of ICASSP}, pp. 5505--5509, 2016.

\bibitem{oord-arXiv-2016}
A.~v.~d. Oord, S.~Dieleman, H.~Zen, K.~Simonyan, O.~Vinyals, A.~Graves,
  N.~Kalchbrenner, A.~W. Senior, and K.~Kavukcuoglu,
\newblock ``{W}ave{N}et: {A} generative model for raw audio,''
\newblock {\em CoRR}, vol. abs/1609.03499, 2016.

\bibitem{mehri-arXiv-2016}
S.~Mehri, K.~Kumar, I.~Gulrajani, R.~Kumar, S.~Jain, J.~Sotelo, A.~Courville,
  and Y.~Bengio,
\newblock ``{S}ample{RNN}: An unconditional end-to-end neural audio generation
  model,''
\newblock {\em CoRR}, vol. abs/1612.07837, 2016.

\bibitem{kalchbrenner-arXiv-2018}
N.~Kalchbrenner, E.~Elsen, K.~Simonyan, S.~Noury, N.~Casagrande, E.~Lockhart,
  F.~Stimberg, A.~v.~d. Oord, S.~Dieleman, and K.~Kavukcuoglu,
\newblock ``Efficient neural audio synthesis,''
\newblock {\em CoRR}, vol. abs/1802.08435, 2018.

\bibitem{jin-ICASSP-2018}
Z.~Jin, A.~Finkelstein, G.~J. Mysore, and J.~Lu,
\newblock ``{FFT}net: A real-time speaker-dependent neural vocoder,''
\newblock {\em Proceedings of ICASSP}, pp. 2251--2255, 2018.

\bibitem{prenger-arXiv-2018}
R.~Prenger, R.~Valle, and B.~Catanzaro,
\newblock ``{W}ave{G}low: A flow-based generative network for speech
  synthesis,''
\newblock {\em CoRR}, vol. abs/1811.00002v1, 2018.

\bibitem{tamamori-ICASSP-2017}
A.~Tamamori, T.~Hayashi, K.~Kovayashi, K.~Takeda, and T.~Toda,
\newblock ``Speaker-dependent {W}ave{N}et vocoder,''
\newblock {\em Proceedings of ICASSP}, pp. 1118--1122, 2017.

\bibitem{robinson-NIPS-1988}
A.~J. Robinson and F.~Fallside,
\newblock ``Static and dynamic error propagation networks with application to
  speech coding,''
\newblock {\em Proceedings of NIPS}, pp. 632--641, 1988.

\bibitem{hochreiter-NeuralComput-1997}
S.~Hochreiter and J.~Schmidhuber,
\newblock ``Long short-term memory,''
\newblock {\em Neural Comput.}, pp. 1735--1780, 1997.

\bibitem{tokuda-ICASSP-2000}
K.~Tokuda, T.~Yoshimura, T.~Masuko, T.~Kobayashi, and T.~Kitamura,
\newblock ``Speech parameter generation algorithms for {HMM}-based speech
  synthesis,''
\newblock {\em Proceedings of ICASSP}, vol. 3, pp. 1315--1318, 2000.

\bibitem{zen-ICASSP-2015}
H.~Zen and H.~Sak,
\newblock ``Unidirectional long short-term memory recurrent neural network with
  recurrent output layer for low-latency speech synthesis,''
\newblock {\em Proceedings of ICASSP}, pp. 4470--4474, 2015.

\bibitem{wang-ICASSP-2017}
W.~Wang and B.~Xu,
\newblock ``Combining unidirectional long short-term memory with convolutional
  output layer for high-performance speech synthesis,''
\newblock {\em Proceedings of ICASSP}, pp. 5500--5504, 2017.

\bibitem{hono-APSIPA-2018}
Y.~Hono, S.~Murata, K.~Nakamura, K.~Hashimoto, K.~Oura, Y.~Nankaku, and
  K.~Tokuda,
\newblock ``Recent development of the {DNN}-based singing voice synthesis
  system - {Sinsy},''
\newblock {\em Proceedings of APSIPA ASC}, pp. 1003--1009, 2018.

\bibitem{merlijn-arXiv-2017}
M.~Blaauw and J.~Bonada,
\newblock ``A neural parametric singing synthesizer,''
\newblock {\em CoRR}, vol. abs/1704.03809, 2013.

\bibitem{yuan-interspeech-2019}
Y.-H. Yi, Y.~Ai, Z.-H. Ling, and L.-R. Dai,
\newblock ``Singing voice synthesis using deep autoregressive neural networks
  for acoustic modeling,''
\newblock {\em Proceedings of Interspeech}, pp. 2593--2597, 2019.

\bibitem{juheon-interspeech-2019}
J.~Lee, H.-S. Choi, C.-B. Jeon, J.~Koo, and K.~Lee,
\newblock ``Adversarially trained end-to-end {K}orean singing voice synthesis
  system,''
\newblock {\em Proceedings of Interspeech}, pp. 2588--2592, 2019.

\bibitem{hashimoto-ICASSP-2015}
K.~Hashimoto, K.~Oura, Y.~Nankaku, and K.~Tokuda,
\newblock ``The effect of neural networks in statistical parametric speech
  synthesis,''
\newblock {\em Proceedings of ICASSP}, pp. 4455--4459, 2015.

\bibitem{imai-IEICE-1983}
S.~Imai, K.~Sumita, and C.~Furuichi,
\newblock ``Mel log spectral approximation filter for speech synthesis,''
\newblock {\em {IECE} Translations on Fundamentals (Japanese Edition)}, vol.
  J66-A, pp. 122--129, 1983.

\bibitem{zen-SCommu-2009}
H.~Zen, K.~Tokuda, and A.~W. Black,
\newblock ``Statistical parametric speech synthesis,''
\newblock {\em Speech Communication}, vol. 51, no. 11, pp. 1039--1064, 2009.

\bibitem{fan-Interspeech-2014}
Y.~Fan, Y.~Qian, F.~Xie, and F.~K. Soong,
\newblock ``{TTS} synthesis with bidirectional {LSTM} based recurrent neural
  networks,''
\newblock {\em Proceeding of Interspeech}, pp. 964--1968, 2014.

\bibitem{fernandez-Interspeech-2014}
R.~Fernandez, A.~Rendel, B.~Ramabhadren, and R.~Hoory,
\newblock ``Prosody contour prediction with long short-term memory,
  bidirectional, deep recurrent neural networks,''
\newblock {\em Proceedings of Interspeech}, pp. 2268--2272, 2014.

\bibitem{hashimoto-ICASSP-2016}
K.~Hashimoto, K.~Oura, Y.~Nankaku, and K.~Tokuda,
\newblock ``Trajectory training considering global variance for speech
  synthesis based on neural networks,''
\newblock {\em Proceedings of ICASSP}, pp. 5600--5604, 2016.

\bibitem{long-CVPR-2015}
J.~Long, E.~Shelhamer, and T.~Darrell,
\newblock ``Fully convolutional networks for semantic segmentation,''
\newblock {\em Proceedings of CVPR}, pp. 3431--3440, 2015.

\bibitem{shen-icassp-2018}
J.~Shen, R.~Pang, R.~J. Weiss, M.~Schuster, N.~Jaitly, Z.~Yang, Z.~Chen,
  Y.~Zhang, Y.~Wang, R.~Skerry-Ryan, R.~A. Saurous, Y.~Agiomyrgiannakis, and
  Y.~Wu,
\newblock ``Natural {TTS} synthesis by conditioning {W}ave{N}et on {M}el
  spectrogram predictions,''
\newblock {\em Proceedings of ICASSP}, pp. 4779--4783, 2018.

\bibitem{li-arXiv-2018}
N.~Li, S.~Liu, Y.~Liu, S.~Zhao, M.~Liu, and M.~Zhou,
\newblock ``Close to human quality {TTS} with transformer,''
\newblock {\em CoRR}, vol. abs/1809.08895, 2018.

\bibitem{kawahara-SpeechCommunication-1999}
H.~Kawahara, M.~K. Ikuyo, and A.~d.~Cheveigne,
\newblock ``Restructuring speech representations using the pitch-adaptive
  time-frequency smoothing and an instantaneous-frequency-based f0 extraction:
  Possible role of a repetitive structure in sounds,''
\newblock {\em Speech Communication}, vol. 27, no. 3, pp. 187--207, 1999.

\bibitem{PCM}
``Pulse code modulation ({PCM}) of voice frequencies,''
\newblock {\em ITU-T Recommendation G.711}, 1988.

\bibitem{yoshimura-IEEE-2018}
T.~Yoshimura, K.~Hashimoto, K.~Oura, Y.~Nankaku, and K.~Tokuda,
\newblock ``Mel-cepstrum-based quantization noise shaping applied to
  neural-network-based speech waveform synthesis,''
\newblock {\em IEEE/ACM Transactions on Audio, Speech, and Language
  Processing}, vol. 26, no. 7, pp. 1173--1180, 2018.

\bibitem{techno-speech-web}
{Techno-Speech,Inc.},
\newblock ``Reproducing high-quality singing voice with state-of-the-art {AI}
  technology,''
\newblock 2018,
\newblock \url{https://www.techno-speech.com/news-20181214a-en}.

\end{thebibliography}

\end{document}